\documentclass[%
 reprint,
superscriptaddress,
 amsmath,amssymb,
]{revtex4-1}
\usepackage{graphicx}% Include figure files
\usepackage{dcolumn}% Align table columns on decimal point
\usepackage{bm}% bold math
\usepackage{xcolor}
\usepackage{siunitx}
\begin{document}

%\preprint{APS/123-QED}

\title{Experimental generation of circulating cavity magnon polaritons}% Force line breaks with \\

\author{Jeremy Bourhill}
\email{jeremy.bourhill@uwa.edu.au}
  \affiliation{ARC Centre of Excellence for Engineered Quantum Systems and ARC Centre of Excellence for Dark Matter
Particle Physics, Department of Physics, University of Western Australia, 35 Stirling Highway, Crawley, Western
Australia 6009, Australia}
\author{Weichao Yu}
 \affiliation{State Key Laboratory of Surface Physics and Institute for Nanoelectronic Devices and Quantum Computing, Fudan University, Shanghai, 200433 China}
 \affiliation{Zhangjiang Fudan International Innovation Center, Fudan University, Shanghai, 201210 China}
 \author{Vincent Vlaminck}
  \affiliation{IMT Atlantique, Technopole Brest-Iroise, CS 83818, 29238 Brest Cedex 3, France}
 \affiliation{Lab-STICC (UMR 6285), CNRS, Technopole Brest-Iroise, CS 83818, 29238 Brest Cedex 3, France}%
 \author{Gerrit E. W. Bauer}
\affiliation{WPI-AIMR and IMR, Tohoku University, Sendai 980-8577, Japan}
\affiliation{Kavli Institute for Theoretical Sciences, University of the Chinese Academy of Sciences, Beijing 10090, China}
\affiliation{Zernike Institute for Advanced Materials, Groningen University, Groningen, Netherlands}
\affiliation{Kavli Institute for Theoretical Sciences, University of the Chinese Acadamy of Sciences, Beijing, China}
 \author{Giuseppe Ruoso}
 \affiliation{INFN, Laboratori Nazionali di Legnaro, Viale Dell{'}Universit\`{a} 2, 35020 Legnaro, Padova, Italy}
 \author{Vincent Castel}
 \affiliation{IMT Atlantique, Technopole Brest-Iroise, CS 83818, 29238 Brest Cedex 3, France}
 \affiliation{Lab-STICC (UMR 6285), CNRS, Technopole Brest-Iroise, CS 83818, 29238 Brest Cedex 3, France}%

\date{\today}% It is always \today, today,

\begin{abstract}
We experimentally realize circularly polarised unidirectional cavity magnon
polaritons in a torus-shaped microwave cavity loaded by a small magnetic
sphere. At special positions the clockwise and counterclockwise modes are
circularly polarized, such that only one of them couples to the magnet,
which breaks the mode degeneracy. We reveal the chiral nature of the
spectral energy and angular momentum flow by observing and modelling
non-reciprocities of the microwave scattering matrix.
\end{abstract}

\maketitle

\section{Introduction}

The study of cavity magnonics lead to the discovery of new and interesting
regimes of the coupling dynamics between photons and magnons. The field
attracted initial attention by the relative ease of achieving strong
coupling, even at room temperature \cite{Tabuchi113,Huebl,Zhang1} because of
the large spin density in magnetic materials leads to a large coupling that
easily exceeds cavity loss rates and thereby hybridized quasi-particles
(magnon polaritons). Attractive is also the tunability of the magnetic
resonance by an applied external DC magnetic field through the cavity modes.

Specially designed cavities \cite{goryachev14,Flower19} and larger magnetic
samples \cite{Zhang1,Bourhill16} reach the ultra-strong coupling regime, in
which the magnon-photon coupling rate is a sizeable fraction of the system
frequency. Dissipative coupling \cite{Harder,NatCom8,Rao:2021aa} of magnets
in open or leaky waveguides strongly modifies the dynamics, e.g. causing
resonance level attraction. Multiple magnets inside a cavity, also called
magnonic molecules, \cite%
{crescini2020coherent,PhysRevA.93.021803,PhysRevB.100.094415,Rameshti,Zhang15}
sustain delocalized collective modes by the real or virtual exchange of
cavity photons.

The observed non-reciprocity in (open or closed) microwave cavities is
interesting for applications since nonreciprocal microwave components are
essential elements in classical and possibly quantum communication systems 
\cite{tang,PhysRevApplied.10.047001,doi:10.1063/5.0063786}. An isolator is a
diode-like device that blocks signal flow in one direction, thereby
protecting a circuit from unintended reflections, e.g. isolating the
transmitter from the receiver in a radar architecture \cite{radar,4534576},
to shield qubits from their environment \cite%
{Abdo:2019aa,PhysRevLett.106.110502,doi:10.1126/science.1226897}, or to
investigate topological materials \cite{Huang_2022,PhysRevB.100.115431}. Yu
et al. \cite{weichao} proposed generating non-reciprocity in a circular
microwave cavity by breaking its time-reversal symmetry (TRS) with magnetic
loads with preferential clockwise vs. counterclockwise spin and energy
flows. The TRS breaking can be achieved by positioning magnets on special
chiral lines in a microwave cavity on which the propagating photons are
chiral \cite{PhysRevB.101.094414,PhysRevLett.124.107202}, i.e. the sign of
their circular polarization is locked to their linear momentum. This causes
a strong direction-dependent coupling with the magnon excitations that can
be controlled by applied magnetic fields.

Here, we experimentally demonstrate circularly polarized, unidirectional
magnon polaritons, thereby confirming theoretical predictions \cite{weichao}%
. We place a YIG sphere inside a newly machined torus-shaped cavity{\ on
special positions and tune the magnetic resonance to a } transverse electric
(TE) \ cavity mode. We detect the coupling dynamics in the microwave
scattering matrix as a function of an applied magnetic field. The
experimental results agree with the simulations and the non-reciprocity of
scattering parameters confirm the chiral nature of the hybrid modes. 

\begin{figure}[t]
\includegraphics[width=7cm]{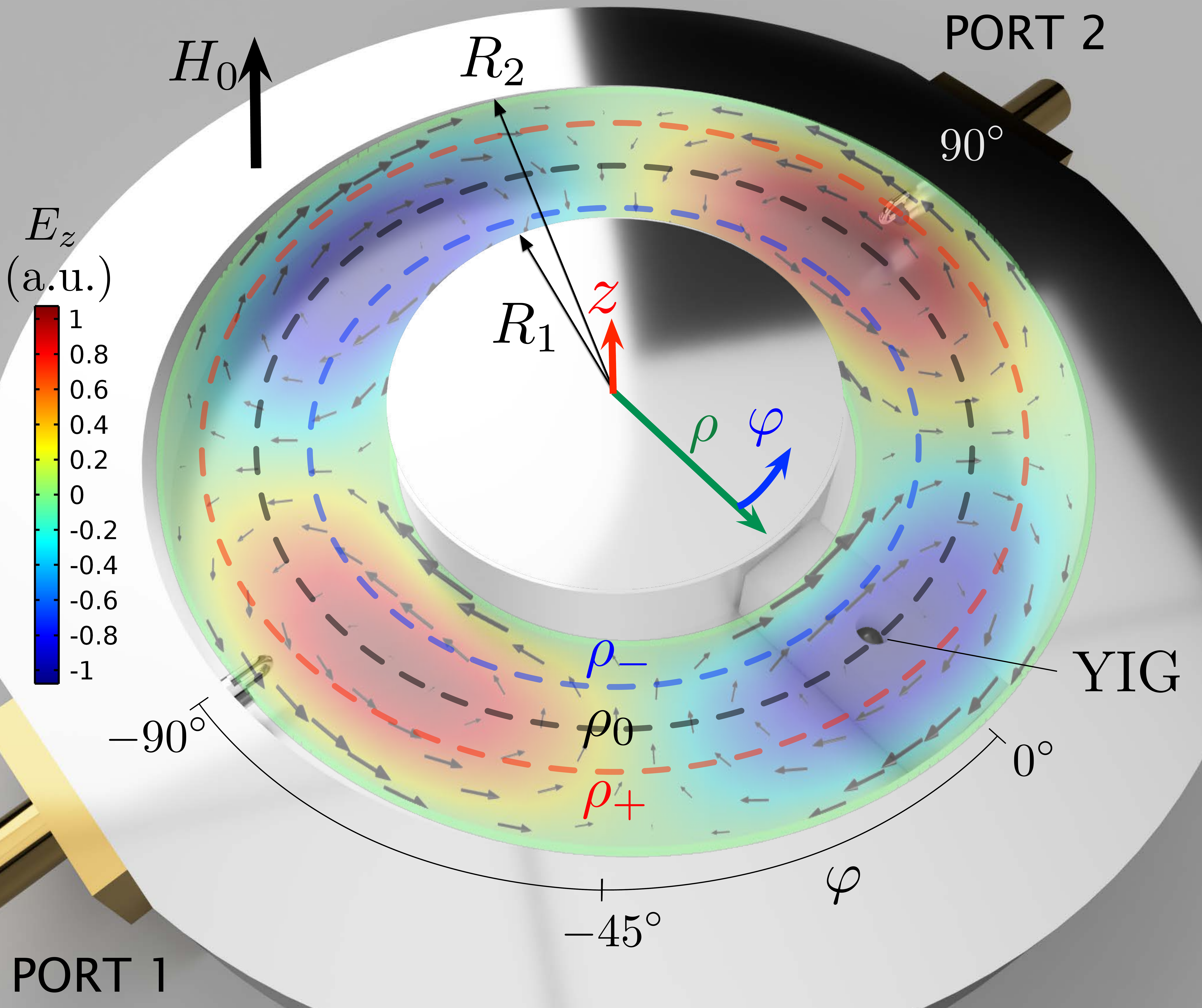}
\caption{Torus-shaped Al cavity overlayed with a snapshot of the simulated
electromagnetic fields of the $m=2$ TE cavity mode. The $E_{z}$ amplitude is
coded by the color density plot, while the arrow vector plot represents $%
\vec{H}$. The dashed lines indicate the special radial positions $\protect%
\rho_{-}$, $\protect\rho_{0}$ and $\protect\rho_{+}$ (see text). The YIG
sphere in the sample holder is the black dot at $\protect\rho=\protect\rho%
_{0}$, $\protect\varphi=0$.}
\label{fig:syst}
\end{figure}

\section{System}

The torus-shaped microwave cavity shown in Fig. 1 is machined from high
purity aluminium in two parts (lid and base). The inner and outer radii $%
R_{1}$ (15 mm) and $R_{2}$ (30 mm) and the height $h$ (6 mm) were chosen to
match the model parameters in Ref. \cite{weichao}. Rather than drive the
cavity by electrical probes in the $z$-axis \cite{weichao}, we place two
magnetic loops around the radial axis at $\varphi=\pm~90^{\circ}$ that
couple to the $H_{\phi}$ component of the cavity field as shown in Fig.\ref%
{fig:syst}. These probes do not measurably affect the frequency response or
RF field distribution. The second azimuthal harmonic of the TE mode of the
torus cavity resonates at $f_{c}=10.8$ GHz with a quality factor $Q=775$,
which corresponds to a bandwidth of $\kappa /2\pi=14$ MHz.

%The configuration of the probes chosen allowed the distance between the poles of the electromagnet to be reduced and ensured a uniform DC magnetic field across the magnetic sample placed inside the cavity.

%Even with a cavity polishing step (mirroring), the expected Q-factor estimated by CST simulation ($Q=2400$) cannot be achieved, mainly due to the residual surface roughness and the two-part cavity configuration, which reduced the effective conductivity at their contact surface. The CST simulations assumed a perfect contact between base and lid without any roughness.

The pure single-crystal YIG sphere with a diameter of 1.8 mm was provided by
INFN-Laboratori Nazionali di Legnaro. The grinding procedure \cite%
{crescini2020coherent} led to a ferromagnetic resonance linewidth of $\sim $%
4 MHz at 10 GHz. The magnon-photon coupling strength was mapped by putting
the YIG sphere at different positions inside the cavity by means of a sample
holder made from Rohacell\textsuperscript{\textregistered} HF, with a hole
drilled at the correct radial coordinates. All measurements were performed
at room temperature. We align the external DC magnetic field $H_{0}$ along
the easy axis of the magnetic sphere and the $z$-axis of the cavity, i.e. $%
\left\langle 111\right\rangle \parallel z$. Microwave spectra were recorded
using a vector network analyzer at both ports as a function of external DC
magnetic field strength.

\section{Theory}

The torus cavity a rectangular waveguide \cite%
{PhysRevB.101.094414,PhysRevLett.124.107202} bent into a circle with
specific azimuthal lines in the cavity at which the microwave magnetic field
of the TE modes with $H_{z}=0$ and $\partial_{n}E_{z}|_{S}=0$ is circularly
polarized \cite{weichao}. Their radial position follow from the analytic
solution of Maxwell's equations in free space with perfect conducting
boundary conditions at the metallic walls of the torus.

%The TE mode has only $E_z$, $H_\varphi$, $H_\rho$ components. The analytical expression for the $E$-field of a TE mode can be found by solving Maxwell's equations in free space with conducting boundary conditions at the metallic walls:
%	\begin{equation}
%	E_z(\rho,\varphi,z)=R_m(\rho)e^{im\varphi}\cos\left(\frac{p\pi z}{h}\right),
%	\end{equation}
%where the integers $p$ and $m$ are the mode numbers in the $z$ and $\varphi$ directions, respectively, and the radial dependence $R_m(\rho)$ can be solved by taking the general Bessel function solution to the wave equation and setting $E_z=0$ at $\rho=R_1,R_2$ \cite{weichao}. Here, we only deal with $p=0$ modes in which $E_z$ is homogenous along the $z$ direction.
%%	\begin{equation}
%%	R_m(\rho)=J_m(\gamma_{p,m}R_1)-\frac{J_m(\gamma_{p,m}R_1)}{N_m(\gamma_{p,m}R_1)}N_m(\gamma_{p,m}\rho),
%%	\end{equation}
%%where $J_m$ and $N_m$ are the Bessel functions of the first and second kind, respectively, and $\gamma_{p,m}$ can be found by
%The two $H$ fields can then be written as:
%\begin{align}
%\label{eq:hrho}
%H_\rho(\rho,\varphi)& =\frac{1}{\mu_0\gamma_m c}\frac{m}{\rho}E_z,\\
%H_\varphi(\rho,\varphi)& =-i\frac{1}{\mu_0\gamma_m c}\frac{\partial E_z}{\partial \rho},
%\end{align}
%where $\mu_0$ and $c$ are the permeability and speed of light in vacuum, and $\gamma_m=\omega_c/c$ with frequency $\omega_c$.

Due to the axial symmetry of the empty cavity the field amplitudes are
proportional to $e^{im\varphi},$ where $m$ is an integer that labels the
orbital angular momentum of clockwise (cw, $m>0$) and counterclockwise (ccw, 
$m<0$) photon circulation. Their linear combinations form phase-shifted
standing waves with frequencies that may be split by a perturbation such as
a YIG\ sphere.

The anticlockwise uniform precession of the magnetization (Kittel mode)
around the effective magnetic field couples to photons with the same
polarization \cite{PhysRevLett.125.027201}. Since we align the easy axis and 
$H_{0}$, a circularly polarized RF magnetic field couples to the Kittel mode
only when right-hand circularly polarized ($|R\rangle$) with respect to the $%
z$-axis. For a TE mode, $|R\rangle$ ($|L\rangle$) polarization corresponds
to the relation $H_{\varphi}=iH_{\rho}$ ($H_{\varphi}=-iH_{\rho})$ between
the polar components of the ac magnetic field. These conditions are
fulfilled at specific radial locations $\rho_{\pm}$ for all $m$.

%, given that
%	\begin{equation}
%	\frac{m}{\rho_\pm}E_z(\rho_\pm)+\frac{\partial  E_z(\rho)}{\partial \rho} \bigg\rvert_{\rho=\rho_\pm}=0.
%	\label{eq:condition}
%	\end{equation}
For a cw mode, the magnetic field has $|L\rangle$ polarization at $\rho_{-}$
and $|R\rangle$ at $\rho_{+}$, and \textit{vice versa} for a ccw mode. A
sphere at these locations with magnetization along $z$ couples to only one
of these modes and breaks the degeneracy between them.

\begin{figure}[b]
\includegraphics[width=9cm]{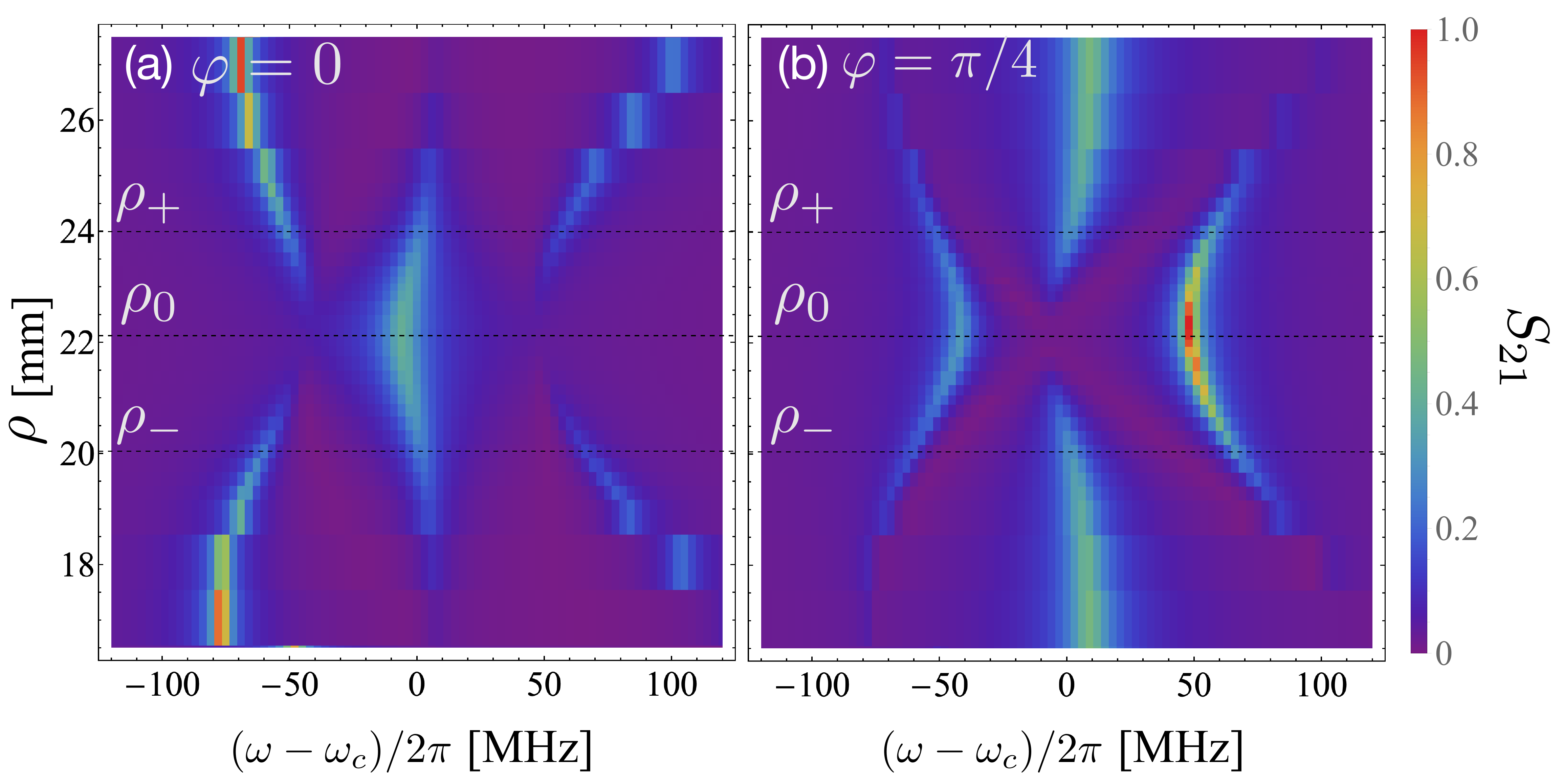}
\caption{Simulated microwave transmission amplitudes for $m=2$ TE mode of
the torus cavity as a function of the radial sphere position along (a) $%
\protect\varphi=0^{\circ}$ and (b) $\protect\varphi=45^{\circ}$ and $%
H_{0}=0.3895$ T tuned such that $\protect\omega_{m}\approx\protect\omega_{c}$%
. Special radial locations are marked by dashed lines.}
\label{fig:0v45}
\end{figure}

\begin{figure*}[t]
\caption{Microwave transmission spectra $\left\vert S_{12}\right\vert $
close to the torus cavity $m=2$ TE mode ($\protect\omega_{c}=10.8$ GHz) and
the Kittel mode of a 1.8 mm diameter YIG sphere located at $\{\protect\varphi%
,\protect\rho\}=$ (a) $\{0,\protect\rho_{-}\}$, (b) $\{0,\protect\rho _{+}\}$%
, (c) $\{0,\protect\rho_{0}\}$, and (d) $\{\protect\pi/4,\protect\rho_{0}\}$%
. The dashed lines are the eigenfrequency solutions of Eq. (\protect\ref%
{eq:ham}) with (a) $g_+=0$, (b) $g_-=0$, (c) $g_+=g_-=J=0$, and (d) is for a
standard two coupled harmonic oscillator model. The panels on the right are
line plots for the resonance $\protect\omega_{m}/\protect\omega_{c}\approx1$%
. }
\label{fig:spectroscopy}\includegraphics[width=18cm]{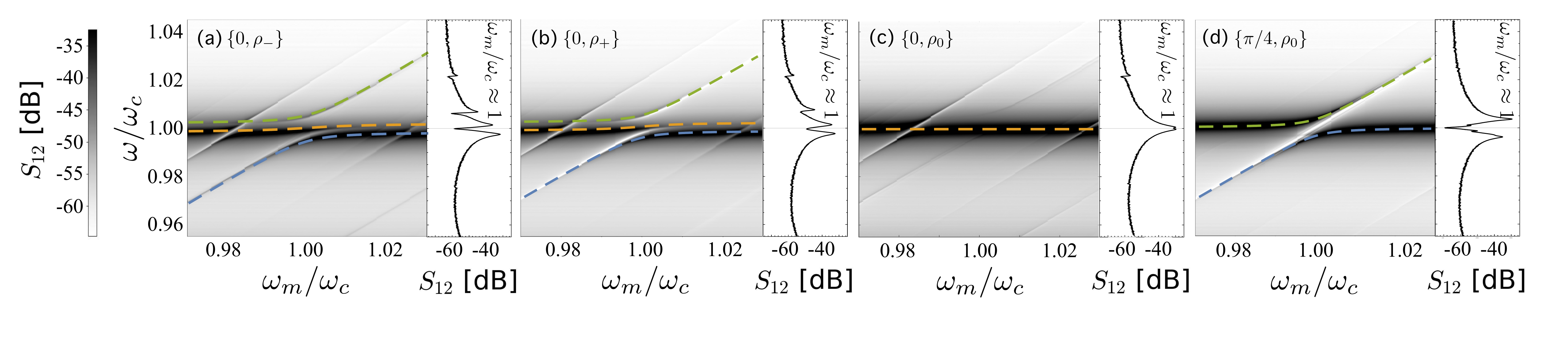}
\par
%\textbf{GB}:
%\textit{In (a) and (b) the green and blue lines do not converge to the orange
%line far from the avoided crossing. The orange line is a bit tilted, so not
%the bare cavity mode. Please explain!}}%
\end{figure*}

We define $\rho_{0}$ to the maximum of the linearly polarized $E_{z}$ field,
while $H_{\varphi}(\rho_{0},\varphi)=0$. Unlike $\rho_{\pm}$, the coupling
along $\rho_{0}$ depends on the angle $\varphi$ of the sphere position. The
magnetic field is then linearly polarized with $H_{\rho}\propto\sin
(m\varphi),$ hence maxima at $\varphi_{1}=(n+1/2)\pi/m$ and zeroes at $%
\varphi_{2}=n\pi/m$ for $n=0,1,2,\ldots$. $\ $At $\varphi_{2}$ cavity and
Kittel modes maximally hybridize. The magnon-photon coupling is therefore
chiral on $\rho=\rho_{\pm},$ not chiral and oscillating at $\rho=\rho_{0}$,
and partially chiral elsewhere \cite{PhysRevB.101.094414}.\newline

We calculate the microwave-induced dynamics of the Kittel mode by solving
the coupled Maxwell and linearized Landau-Lifshitz-Gilbert (LLG) equation in
the frequency domain in the macrospin and rotating-wave approximations \cite%
{Rameshti,PhysRevB.91.094423} by an electromagnetic finite element solver in
the frequency domain. We model the YIG sphere using a permittivity $%
\epsilon_{r}=15$ \cite{doi:10.1142/S0217979209063109} and permeability $%
\mu_{r}$ that takes into account the external magnetic field and frequency
dependent magnetic susceptibility:%
\begin{equation}
\mu_{r}=\begin{pmatrix} 1+u & -iv & 0\\ iv & 1+u & 0\\ 0 & 0 & 1
\end{pmatrix} ,
\end{equation}
where $u=(\omega_{K}-i\alpha\omega)\omega_{M}/[(\omega_{K}-i\alpha\omega
)^{2}-\omega^{2}]$, $v=\omega\omega_{M}/[(\omega_{K}-i\alpha\omega)^{2}-%
\omega^{2}]$, $\omega$ is the microwave frequency, $\omega_{M}=\gamma M_{s}$%
, $\omega_{K}=\gamma H_{0}$ the Kittel mode frequency, $\gamma$ the
gyromagnetic ratio and $\alpha$ the Gilbert damping constant. We the can
calculate the observables, \textit{viz}. the microwave transmission and
reflection spectra as measured by the two ports to the cavity \cite{weichao}

We may infer the position-dependent coupling between the $m=2$ TE cavity and
Kittel mode from the calculated transmission in Fig. \ref{fig:0v45} as a
function of the microwave feed frequency and radial sphere position for two
azimuthal angles. We observe all qualitative features mentioned above, such
as a vanishing coupling at the position $\rho_{0}$ in (a) compared to the
anticrossing in (b). At $\rho_{\pm}$, the interaction splits two modes,
rendering them all visible at any angle.\newline

%For $\rho\neq\rho_0,\rho_\pm$, the above relations still hold, but $H_\varphi\neq0$, and the magnon-photon coupling is partially chiral \cite{PhysRevB.101.094414}.

The system can be accurately modelled by three coupled harmonic oscillators,
viz. the two counter-rotating photonic modes and the magnetic mode. The
second-quantized Hamiltonian reads 
\begin{equation}
\begin{aligned} H_\text{sys}=&~\hbar \omega_c a^\dagger a+\hbar \omega_c
b^\dagger b+\hbar \omega_m m^\dagger m\\ &+\hbar g_{+}(a^\dagger m
+m^\dagger a)+\hbar g_{-}(b^\dagger m +m^\dagger b)\\ &+\hbar J (a^\dagger
b+b^\dagger a) \end{aligned}  \label{eq:ham}
\end{equation}
where $\omega_{c}$, $\omega_{m}$ are the cavity and magnon mode frequencies,
respectively, $a,b,m$ represent annihilation operators for the cw photonic
mode, the ccw photonic mode and the magnon mode, respectively, and the
coupling between the respective photonic modes and the magnon $g_{+}$, $%
g_{-} $\ depend on the position. $J$ \ is a weak coupling between the two
photonic modes by dielectric backscattering from the sphere or by the ports 
\cite{PhysRevB.89.224407}. The action of $J$ is to split the cavity mode
into a doublet even far from the avoided crossing, which is observed
experimentally (see Fig. \ref{fig:spectroscopy}(a) and (b)). At very large $%
H_0$ values, the cavity mode will cease appearing as a doublet, but within
the $\omega_m/\omega_c$ range of Fig. \ref{fig:spectroscopy}, Eq. (\ref%
{eq:ham}) is valid.

Since the microwave field is circularly polarized at the special radial
locations, $g_{+}(\rho_{-},\varphi)=0$ and $g_{-}(\rho_{+},\varphi)=0$ for $%
H_{0}>0$. $g_{+}$ and $g_{-}$ can be expressed analytically \cite{weichao}
in the limit of small sphere radii and calculated numerically by the finite
element method \cite{Bourhill2020}.

At the special radii, our model splits into two separate systems: an
uncoupled magnon mode and two coupled harmonic oscillators that form a
magnon polariton. The amplitudes can be obtained by the standard
input/output formalism for two ports \cite{doi:10.1063/5.0063786} that leads
to the $2\times 2$ scattering matrix $S$. For example, the cw mode interacts
with a sphere at $\rho _{+}$ when $H_{0}>0$ with scattering matrix%
\begin{equation}
\begin{aligned} S=&D_\text{ccw}\left[-A_\text{ccw}-i\omega I
\right]^{-1}B_\text{ccw}\\ &+\zeta D_\text{cw}\left[-A_\text{cw}-i\omega
I\right]^{-1}B_\text{cw}, \end{aligned}  \label{eq:Smatrix}
\end{equation}%
where 
\begin{equation}
\begin{aligned} A_\text{cw}=& \begin{pmatrix} -i
\omega_\text{cw}-\frac{\kappa_0}{2}&i g\\ i g&-i\omega_m-\frac{\kappa_m}{2}
\end{pmatrix},\\ B_\text{cw}=&\sqrt{\frac{\kappa_\text{cw}}{2}}
\begin{pmatrix} 1&e^{i\alpha_\text{cw}}\\ 0&0\\ \end{pmatrix},\\
A_\text{ccw}=&-i
\omega_\text{ccw}-\frac{\kappa_0}{2},~B_\text{ccw}=~\sqrt{\frac{\kappa_%
\text{ccw}}{2}} \begin{pmatrix} 1&e^{i\alpha_\text{ccw}} \end{pmatrix},\\
D_{X}=&-CB_{X}^\dagger,~C=\begin{pmatrix} \sqrt{1-\xi}&\i \sqrt{\xi}\\
i\sqrt{\xi}&\sqrt{1-\xi} \end{pmatrix}, \\ \zeta=&\begin{pmatrix} e^{i\beta}
& 0\\ 0& e^{-i\beta} \end{pmatrix}. \end{aligned}
\end{equation}%
Here, $\omega _{\mathrm{ccw/cw}}$ are the mode frequencies, $\kappa _{0}$
the cavity loss rate, $\kappa _{\mathrm{cw/ccw}}$ the coupling loss rates
assumed equal at port 1 and 2, $\alpha _{\mathrm{cw/ccw}}$ the phase
difference between a mode excited at port 1 or port 2, and $0\leq \xi \leq 1$
represents the crosstalk between the two ports, and we assume $\xi =0$,
meaning no crosstalk. Since the cavity field at different positions is
always in phase for standing waves, we have $\alpha _{X}=0$ \cite%
{doi:10.1063/5.0063786}. We assume that a weak mode coupling $J$ in Eq. (\ref%
{eq:ham}) can be handled by slight shifts of the otherwise degenerate $%
\omega _{\mathrm{cw/ccw}}$ on the order of 20 MHz. $\beta $ is the
propagation phase difference between the two modes that are out-coupled
through the same port. This phase factor depends on the flow direction,
because a cw signal from port 2 first travels through the lower arm of the
cavity with the YIG sample, while that from port 1 first has to travel
through the upper arm before passing at port 2 and arriving at the YIG
sphere because of the unidirectional nature of the propagating signals
introduced by the YIG sphere that is maximized when on the special $\rho
_{\pm }$. In a partially open cavity, these phase differences lead to
signatures of the unidirectional circulation of microwaves in the form of
non-reciprocal transmission $S_{12}\neq S_{21}.$ The couplings $\kappa _{%
\mathrm{cw/ccw}}$ may not be too small; when the photon lifetime in the
cavity is long, the phase accumulated between port and magnet insertion does
not play a role, and the transmission is always reciprocal.

When $\omega_{m}\approx\omega_{c}$, the YIG sphere is at a chiral position,
and the magnon-phonon coupling is much larger than the back scattering $J$,
the transmission spectra are characterized by a single magnon polariton with
two resonances (peaks) at the eigenfrequencies of the hybrid system and an
antiresonance (dip) at the uncoupled frequency\ \cite%
{HARDER201847,PhysRevB.94.054403} and a non-interacting cavity mode. Only
the first term in Eq. (\ref{eq:Smatrix}) depends on the magnetic field and
can lead to non-reciprocity.

We compare the calculated results with the experiments in the following
section, where appropriate.

\begin{figure}[t]
\vspace{0.5cm} \includegraphics[width=8cm]{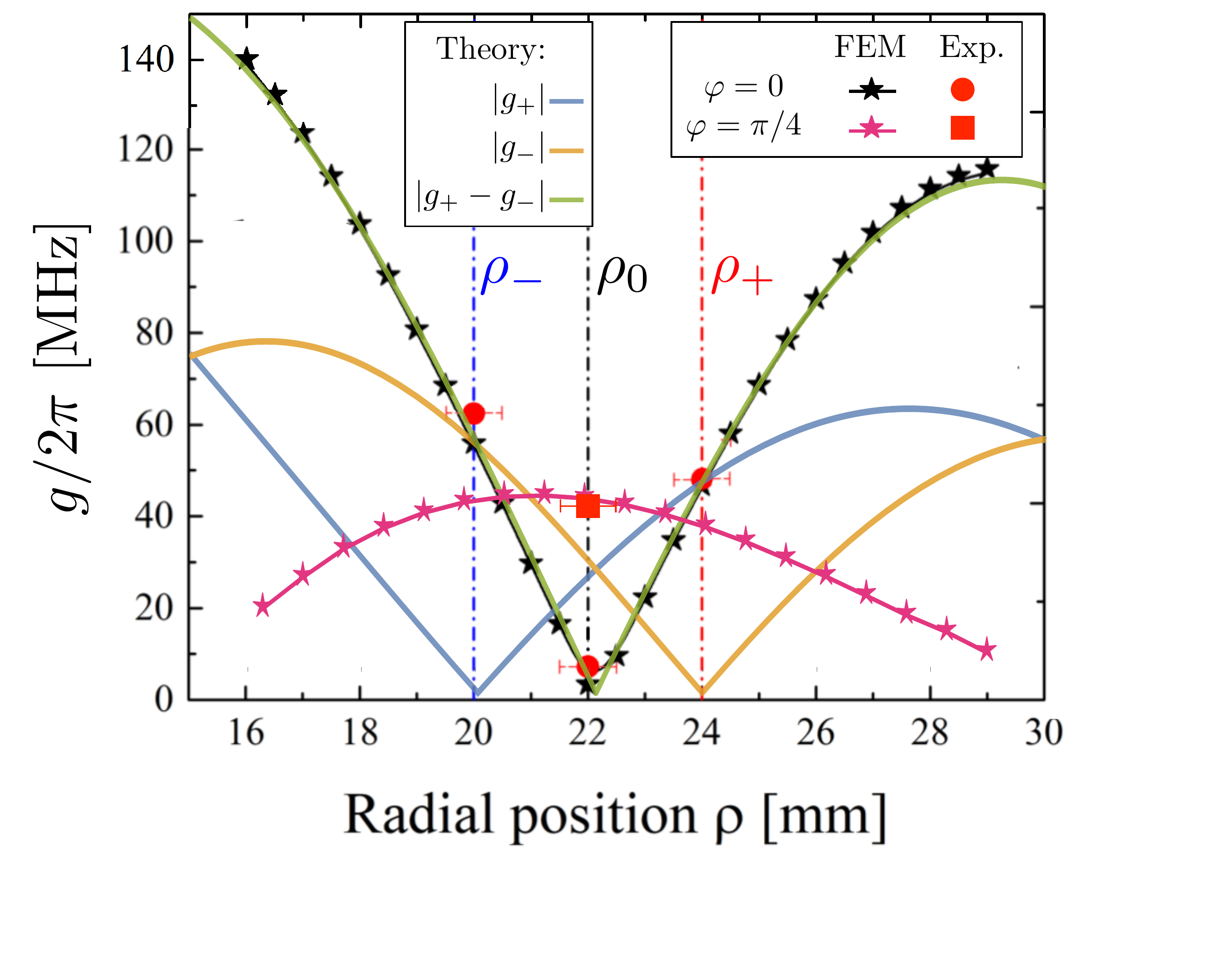}
\caption{Calculated (stars) and measured (red dots and square) magnon-photon
coupling rates with the Kittel mode as a function of radial position $%
\protect\rho$ along $\protect\varphi=0$ and $\protect\varphi=\protect\pi/4$.
The special positions $\protect\rho_{-},\protect\rho_{0}$ and $\protect\rho%
_{+}$ are shown as blue, black and red dashed lines, respectively.}
\label{fig:eta}
\end{figure}

\section{Experimental Results}

\label{sec:exp}

Fig. \ref{fig:spectroscopy} summarizes the microwave transmission spectra $%
S_{21}$ for a 1.8 mm diameter YIG sphere located at different positions
within the torus close to the resonance $\omega_{m}=\omega_{c}$. We clearly
observe a the expected feature of a magnon polariton when placing the YIG
sphere on $\rho_{+}$ and $\rho_{-}$. On the other hand, the coupling nearly
vanishes at $\rho_{0}$ and $\varphi=0^{\circ}$, and changes into a two-mode
avoided level crossing for $\varphi=45^{\circ}$ as predicted \cite{weichao}.

Since the coupling is strong, we can directly read-off the cavity-magnon
coupling rates $g_{\pm}/2\pi$ in Eq. (\ref{eq:ham}) from the experimental
data. The dashed lines in Fig. \ref{fig:spectroscopy} are eigenfrequencies
of (\ref{eq:ham}) with $g_X/2\pi=63$, 48 and 40 MHz for $\{\varphi
,~\rho\}=\{0^{\circ},\rho_{-}\},~\{0^{\circ},~\rho_{+}\}$ and $\{45^{\circ
},~\rho_{0}\}$, respectively. We deduce a much smaller $g_X/2\pi\approx8.5$
MHz for $\{0^{\circ},~\rho_{0}\}$ from the splitting of the line-plot of the 
$S_{12}$ transmission spectra for $\omega_{m}=\omega_{c}$ in the right panel
of Fig.\ref{fig:spectroscopy} (c).

%The three-mode signature of Fig. \ref{fig:spectroscopy} (a) and (b) is a result of the angular momentum selectivity of the magnon-photon interaction; a cw travelling photon will interact and a ccw travelling photon will not, or vice-versa, depending on the sphere{'}s location and external magnetic field direction.
%The non-interaction seen in Fig. \ref{fig:spectroscopy} (c) is a result of $\rho_0$ being the location of an RF magnetic field minima, whilst the standard two-mode interaction of (d) is due to a coupling to linearly polarised RF magnetic field at $\varphi=45^\circ$ and hence no selectivity due to angular momentum of the photon needs to be observed.

Fig. \ref{fig:eta} shows the measured coupling rates between the 1.8 mm
diameter YIG sphere and the $m=2$ TE mode for $\varphi=0,$ (red dots) and $%
\pi/4$ (red square) together with theory and simulations. Stars represent\ $%
g $ as calculated by finite element modelling while blue and yellow lines
are analytical results for $g_{\pm m}$, respectively \cite{weichao}. Theory
and experiments, as well as $\left\vert g_{-m}-g_{m}\right\vert $ (green
line) of the analytic and numerical results agree very well. In the regions $%
\rho_{-}<\rho<\rho_{+}$, the contributions of cw and ccw waves to the
coupling partly cancel, but enhance each other otherwise..

A higher order magnetostatic mode \cite{PhysRevB.97.214423} in Fig. \ref%
{fig:spectroscopy} appears under the main $\omega _{m}=\gamma H_{0}$ line
(Kittel mode). It becomes observable by the non-uniformity of the magnetic
field, which is larger along $\varphi =0^{\circ }$ as compared to $\varphi
=45^{\circ }$, as seen by inspection of Fig. \ref{fig:syst}. This explains
the observed stronger coupling of the higher spin wave for $\varphi
=0^{\circ }$, i.e. Fig. \ref{fig:spectroscopy}(a)-(c) vs. (d). The higher
order (Walker) modes remain observable even when the sphere is at $\{0,\rho
_{0}\}$ because of its finite size which samples ac magnetic field
gradients. The main higher order mode is the same at all $\rho $ at $\omega
/\omega _{c}\approx 1.02$ in Fig. \ref{fig:spectroscopy}. Crossing the
cavity frequency at $\omega _{m}=0.98~\omega _{c}$ it is most likely the $%
\{200\}$ magnetostatic mode \cite{PhysRevB.97.214423}, and appears to be
well described by a weaker two-level avoided crossing, but without the
angular momentum selection rules that govern the Kittel mode resonance.

%\textbf{GB}: \textit{So what mode could this be?
%James Haigh has labeled the modes in the FMR of spheres in }%
%https://journals.aps.org/prb/abstract/10.1103/PhysRevB.97.214423 ,

%given higher order mode{'}s ability to couple to both $|L\rangle$ and $|R\rangle$ polarisations \cite{PhysRevB.97.214423}.

%\subsection{Analysis of transmission phase}
\begin{figure}[t]
\includegraphics[width=8.7cm]{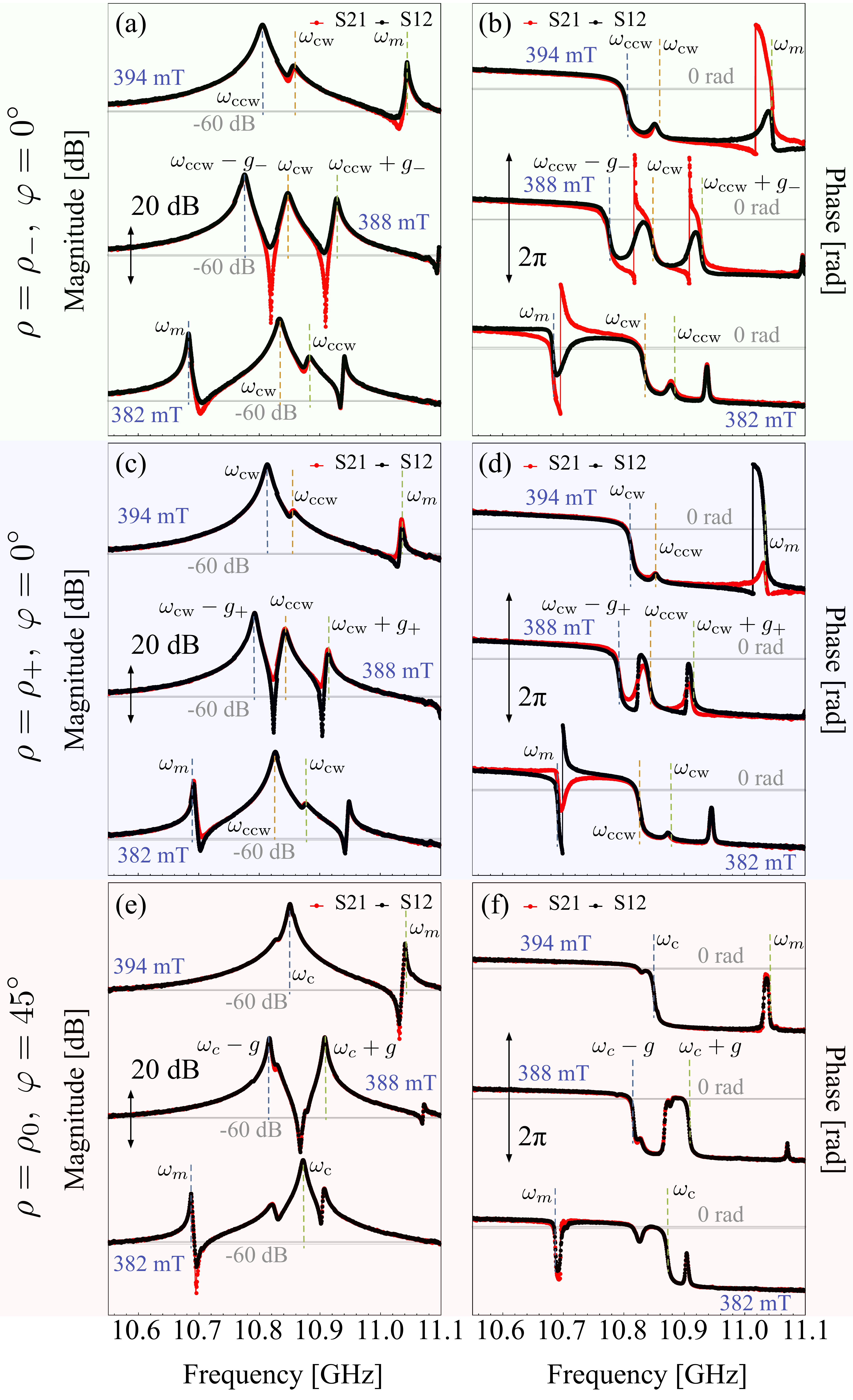}
\caption{Measured modulus (left) and phase (right) of the transmission
amplitudes through a torus cavity with two ports in both directions $S_{21}$
(red line) and $S_{12}$ (black line), loaded by a YIG sphere at $\protect\rho%
_{-}$ (top), $\protect\rho_{+}$ (middle) and $\protect\rho_{0}$, $\protect%
\varphi=\protect\pi/4$ as a function of the applied magnetic field. }
\label{fig:phase}
\end{figure}

A strongly coupled ferromagnetic resonance with a cavity mode generates two
hybrid modes. When crossing the resonances with increasing frequency the
transmission amplitude suffers a $-\pi$ phase shift. The anti-resonance in
the avoided crossing gap is characterized by a dip in the transmission
amplitude and a $+\pi$ phase shift \cite{PhysRevB.94.054403,HARDER201847}.
The unperturbed magnetic mode frequency of an uncoupled cavity mode, on the
other hand, causes a single amplitude peak and a pair of $\left(
\pi,-\pi\right) $ phase shifts. The spectra in Fig. \ref{fig:phase} (a)-(d)
in which the spheres are on the two chiral lines show significant
non-reciprocities, but obey approximately the symmetry relation $%
S_{12}\left( \rho_{+}\right) =$ $S_{21}\left( \rho_{-}\right) $ and \textit{%
vice versa} (this symmetry is not exact, because the couplings are not
exactly the same). Alternatively, a reversal of the external magnetic field
will have the same result (not shown). The significant $S_{12}\left(
\rho_{\pm}\right) \neq$ $S_{21}\left( \rho_{\pm}\right) $ is a strong
evidence of chiral microwave energy flow that is rendered observable by the
breaking of the up-down mirror symmetry by the load. When the magnet is
placed on a non-chiral position at $\{\pi/4,\rho_{0}\}$ we recover a
response as seen in conventional cavities (see Fig. \ref{fig:phase} (e) and
(f)). As expected, we do not observe a significant non-reciprocity in this
case.

The action of the back scattering term $J$ in Eq. (\ref{eq:ham}) is to
couple the non-interacting mode to the interacting mode. Therefore the large
shift in mode frequency caused by the magnon interacting with the latter
creates a smaller knock-on effect in the former, as can be seen by the
slight perturbation of the yellow line in Fig. \ref{fig:spectroscopy}(a) and
(b). This is a detail that is lost in the simplified approximate model of
Eq. (\ref{eq:Smatrix}), which accounts for this effect by defining two mode
frequencies independent of one another, which will be individually dependent
on the applied magnetic field. This is why the frequencies $\omega _{\mathrm{%
cw/ccw}}$ are observed to shift as a function of $H_{0}$ in Fig. \ref%
{fig:phase}.

\begin{figure}[b]
\includegraphics[width=9cm]{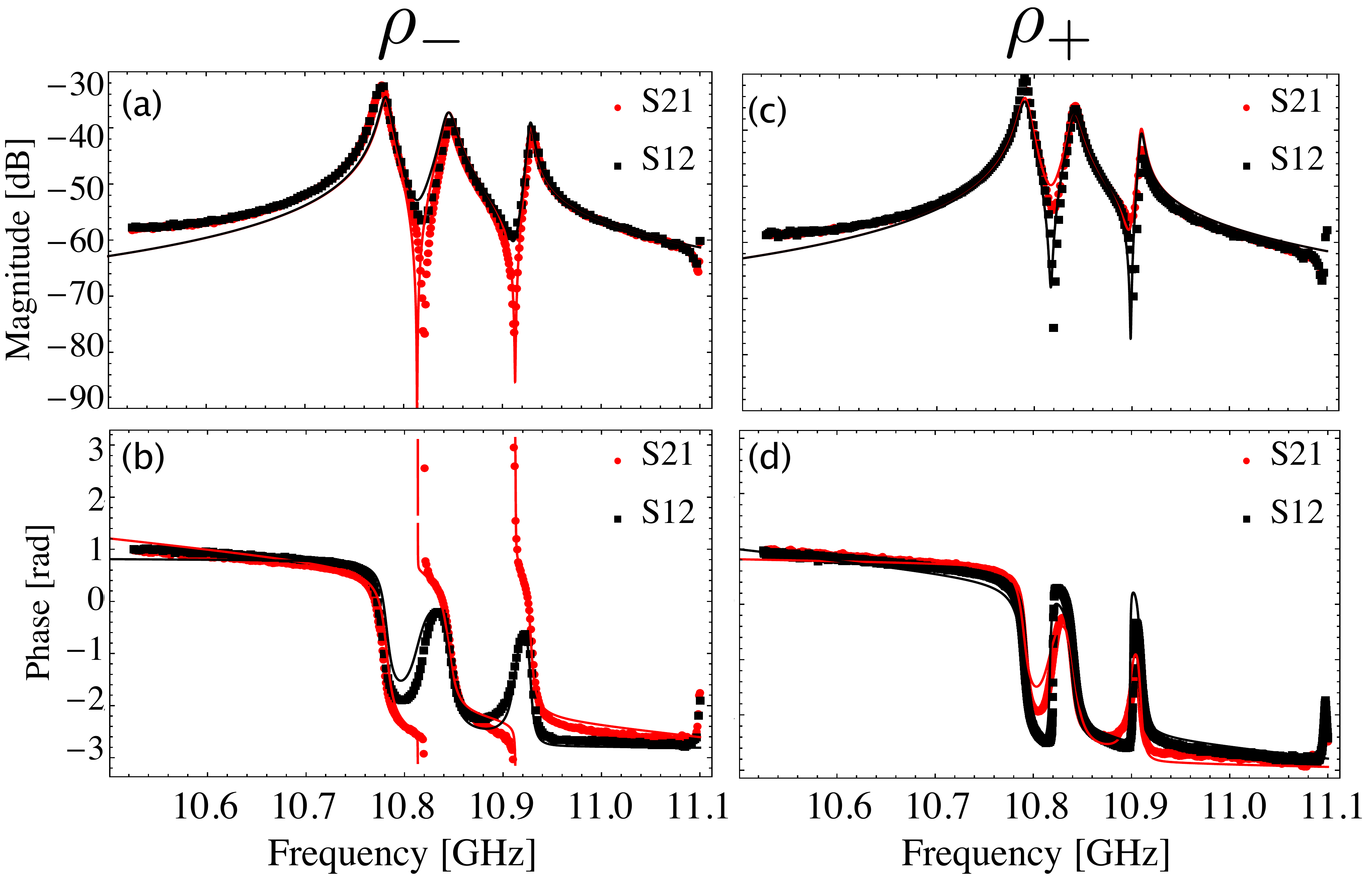}
\caption{Transmission spectra from Fig. 5(a)-(d) for $\protect\omega%
_{m}\approx\protect\omega_{c}$ when the YIG sphere is located at $\protect%
\rho_{-}$ and $\protect\rho_{+}.$ The thin red and black lines are the fit
obtained by the model scattering matrix (\protect\ref{eq:Smatrix}).}
\label{fig:fit}
\end{figure}

While the presence of $J$ complicated the analysis, we can use the
simplified equation (\ref{eq:Smatrix}) at the avoided crossing when $\omega
_{m}\approx \omega _{\mathrm{ccw}}$ and $\left\vert g\right\vert \gg
\left\vert J\right\vert $ to fit the scattering amplitude for the sphere at $%
\rho _{+}$ (for $\rho _{-}$ we have to change the roles of cw and ccw
modes). We can use independently observed values of $H=0.3895$ T, $\kappa
_{0}/2\pi =14$ MHz, $\kappa _{m}/2\pi =3.937$ MHz, $g/2\pi =63$ MHz for $%
\rho _{-}$ and 48 MHz for $\rho _{+}$, and set $\alpha _{\mathrm{cw}}=\alpha
_{\mathrm{ccw}}=0$. The external port coupling rates $\kappa _{\mathrm{cw/ccw%
}}$, $\beta $, and $\omega _{\mathrm{cw/ccw}}$ are fitting parameters. We
model the resonance phase shift $\beta =B~\mathrm{Erf}\left[ (\omega -\omega
_{c})/\kappa _{0}\right] \ $with $B\approx -0.4$, which fits the results for
both $\rho _{+}$ and $\rho _{-}$, with the other parameters set to $\omega _{%
\mathrm{cw}}/2\pi =10.818$ GHz, $\omega _{\mathrm{ccw}}/2\pi =10.845$ GHz, $%
\kappa _{\mathrm{cw}}=0.077\kappa _{0}$, and $\kappa _{\mathrm{ccw}%
}=0.027\kappa _{0}$ for the sphere located at $\rho _{+}$. The $\rho _{-}$
case can be modelled by swapping cw terms with ccw terms in Eq. (\ref%
{eq:Smatrix}) and changing the sign of $B$ because transmission paths $%
S_{21} $ and $S_{12}$ have been effectively swapped. Hence given all other
parameters are equal in amplitude, the difference in the phase shifts
observable for the two radial positions is caused by the slightly stronger
coupling at $\rho _{-}$ \cite{weichao}. Fig. \ref{fig:fit} shows excellent
agreement of the model with experiments for both spectral features and the
non-reciprocity.

\begin{figure}[b]
\includegraphics[width=8.5 cm]{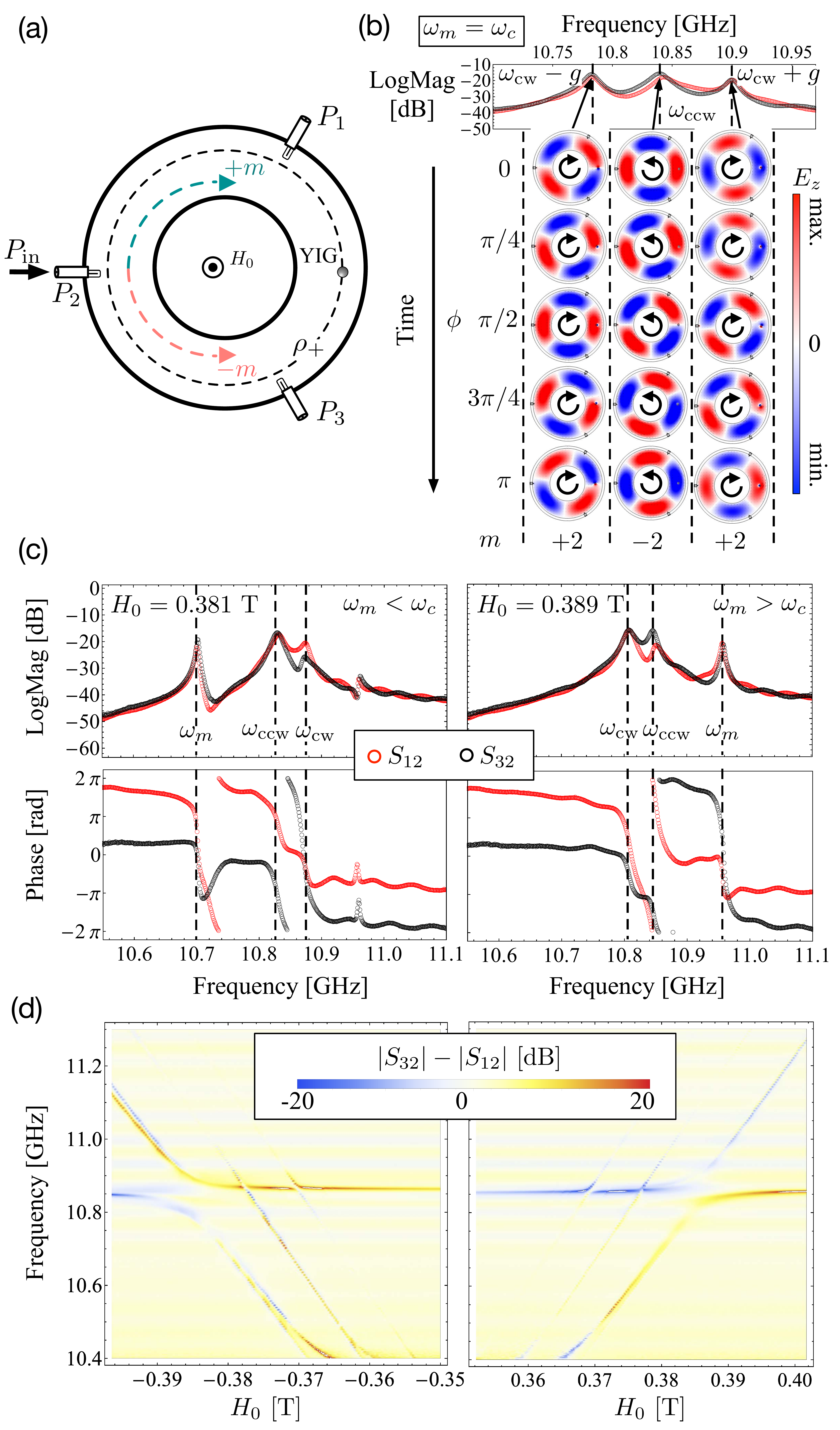}
\caption{(a) Schematic of the 3-port microwave cavity and the position $%
\protect\rho_{+}$ of the magnetic sample. (b) Time evolution of the phase ($%
\protect\phi$) of the eigenmodes calculated at $\protect\omega_{m}=\protect%
\omega_{c}.$ The upper panel shows $|S_{12}|$ in red and $|S_{32}|$ in
black. The lower panel are snapshots of the electric field distribution $%
E_{z}$ at the peaks of $\left\vert S_{xx}\right\vert \left( \protect\omega%
\right) $ that we associate with the magnon polariton (left and right, $m=+2$%
) and the non-interacting mode (centre, $m=-2$) . We clearly observe the
counter rotation of the modes and associated energy and angular momentum
flow as function of microwave frequency. (c) Experimental transmission
amplitude with modulus (top) and phase (bottom) in the three--port setup for
the sphere located at $\protect\rho_{+}$, for $\protect\omega_{m}<\protect%
\omega_{c}$ (left) and $\protect\omega _{m}>\protect\omega_{c}$ (right). (d)
Difference between the the output in ports 1 $\left\vert S_{12}\right\vert $
and 3 $\left\vert S_{32}\right\vert $ when fed via port 2.}
\label{fig:3port2}
\end{figure}

We confirm the unidirectionality of the energy and angular momentum flow in
another torus-shaped cavity with 3 ports located symmetrically at angles of $%
120^{\circ}$, as shown in Fig. \ref{fig:3port2}(a). Using port 2 ($P_{2}$)
as input, unidirectionality should lead to different outputs at $P_{3}$ and $%
P_{1}.$ The phases of $S_{32}$and $S_{12}$ in Fig. \ref{fig:3port2}(c) for $%
\omega_{m}<\omega_{c}$ and $\omega_{m}>\omega_{c}$ are very different when
the sphere is located at $\rho_{+}$.

The phase shifts are associated to the different {\textquotedblleft}arms{%
\textquotedblright\ } \cite{doi:10.1063/5.0063786}, i.e. the blue (upper)
and green (lower) curves in Fig. \ref{fig:spectroscopy}(a). When the magnon
is tuned below the cavity frequency we see in Fig. \ref{fig:3port2}(c) that
a large phase shift is associated with the lower frequency magnon polariton
signal $S_{12}$ via the lower arm, while $S_{32}$ it is associated with the
higher frequency hybrid mode close to the cavity frequency via the upper
arm. When $\omega_{m}>\omega_{c}$ the larger phase of $S_{12}$ is locked to
the cavity frequency, but $S_{32}$ follows $\omega_{m}$. The two branches
therefore maintain their phase shifts.

%The transmission directions swap for the $\rho_+$ case, as would be expected, given the interacting signal will have opposite chirality.

Fig. \ref{fig:3port2}(b) shows the simulated time-evolution of the three
eigenmodes when the YIG sphere is at $\rho_{+}$ and $\omega_{m}=\omega_{c}$.
We see that the magnon polariton modes have a cw ($m>0$) orbital angular
momentum, while the central non-interacting mode moves in opposite direction
($m<0$), as expected.

The asymmetry between $S_{12}$ and $S_{32}$ phases is another direct
evidence of unidirectionality, since it implies that the sphere is not at an
equivalent position relative to the input and output probes. The difference
in transmission amplitudes $|S_{32}|-|S_{12}|$ between the two output ports
in Fig \ref{fig:3port2}(d) for $H_{0}<0$ and $H_{0}>0,$ shows that the upper
branch has a higher transmission amplitude to $P_{1}$, while $P_{3}$ sees a
larger amplitude for the lower branch. This situation is reversed with
magnetic field direction. The Kittel mode then precesses in the opposite
direction, coupling to ccw modes with $|L\rangle$ polarization. Moved the
sphere\ from $\rho_{+}$ to $\rho_{-}$ while keeping $H_{0}$ fixed has a
similar effect (not shown)

%The simulation results also demonstrate how the reversal of mode chirality, shown by the Poynting vector, results in coupling or no coupling to the sphere (note the disappearance of $\vec{M}$ for $\omega=\omega_c=\omega_m$), the direction of which is opposite for the $\rho_-$ and $\rho_+$ cases. This is the same for the two-port setup.

%\begin{figure}[b!]
%		\centering
%		\includegraphics[width=8cm]{splitting2.pdf}
%		\caption{(a) Simulated transmission spectra as a function of cavity losses for a 1.8 mm diameter sphere, demonstrating the derivation of the eigenfrequencies in (b). (b) Simulated eigenfrequencies of the torus cavity--YIG system with $\omega_m\approx\omega_c$, the sphere located at $\rho_+$ and $\varphi=0^\circ$ for three different sphere diameters as cavity loss $\kappa$ is varied.}
%		\label{fig:split}
%\end{figure}	

\section{Discussion}

\label{sec:disc}

%When tested with a 0.5 mm diameter YIG sphere, the three-mode splitting signature was not observable at any position within the cavity. This is a result of the reduced cavity-magnon coupling of the smaller sphere ($\eta\propto \sqrt{V_m}$) leading to the mode-splitting being less than the cavity bandwidth.

%As can be seen in Fig. \ref{fig:split}, cavity loss $\kappa$ must be approximately less than $g_{cm}$ (i.e. in the strong-coupling regime) if the three-mode splitting is to be observable. Therefore, for a small sphere with a small $g_{cm}$, a very low value of $\kappa$ is necessary. In the aforementioned case of the 0.5 mm diameter sphere, $g_{cm}/2\pi=8$ MHz was calculated from simulation, which is approximately equal to the bandwidth of the cavity, hence the three-mode splitting was unobservable.\\

The spontaneous breaking of time reversal symmetry is a unique feature of
magnets that affects the interaction with electromagnetic radiation. In
microwave guides and cavities it enables a chiral magnon-photon coupling
because the polarization of the electromagnetic radiation metallic
boundaries is locked to its orbital motion on {\textquotedblleft}chiral{%
\textquotedblright} lines or planes: A magnet placed at these special lines
breaks the degeneracy of counter-propagating waves and results
unidirectional coupling.

Our observation of non-reciprocity in the scattering matrices of both two--
and three--port torus cavities are a direct proof of microwave circulation,
i.e. unidirectional nature of the magnon polariton modes over macroscopic
distances, which is induced by a relatively very small magnetic load. The
experimental results agree very well with finite-difference calculations of
the coupled Landau-Lifshitz and Maxwell equations and can be understood by a
three-mode input-output model \cite{weichao}. The direction of the microwave
circulation can be reversed by tuning the frequency, flipping the
magnetization direction or shifting the position of magnet. A multiple
sphere setup with magnets placed along the same radial line that couple
coherently can generate a high-power unidirectional photon beam with high
coherence and narrow bandwidth.

\medskip

\textbf{Acknowledgements}

This work was funded by the R\'{e}gion Bretagne through the project
OSCAR-SAD18024. This work is also part of the research program supported by
the European Union through the European Regional Development Fund (ERDF) and
by the Ministry of Higher Education and Research, Brittany and Rennes M\'{e}%
tropole through the CPER Project SOPHIE/STIC \& Ondes. W.Y. is supported by
Shanghai Pujiang Program (No. 21PJ1401500) and Shanghai Science and
Technology Committee (No. 21JC1406200) and G.B. by JSPS Kakenhi Grant \#
19H00645.

%{\color{red}
%The {``}left{''} higher order mode is predicted by the simulations (see Fig. \ref{fig:sim_rho}), the others are not. Given the simulation is electromagnetic only, with a programmable $\mu$ tensor to match the behaviour of the Kittel mode in an external magnetic field, this would suggest that the higher order mode predicted is an electromagnetic resonance in which energy is confined to the sphere itself. Higher order magnon modes are not predicted by the simulation model. Therefore, the avoided level crossing observed with the {``}'left{''} higher order mode is a coupling of two electromagnetic modes, not of a magnon mode with an electromagnetic mode, and hence no selectivity of angular momentum is present.}
%\bibliographystyle{plain}
%\bibliographystyle{plain}
\bibliography{biblio_chiral}

\end{document}